\documentclass[aps,prl,twocolumn,amsmath,amssymb]{revtex4}%
\usepackage{graphicx}
\usepackage{dcolumn}
\usepackage{bm}
\usepackage{color}
\usepackage{amsmath}
\usepackage{epstopdf}
\usepackage{fancybox}
\usepackage{amsfonts}
\usepackage{amssymb}%
\usepackage[toc,page]{appendix}
\renewcommand{\cite}[1]{{[}\onlinecite{#1}{]}}

\newcommand{\be}{\begin{equation}}
\newcommand{\e}{\end{equation}}
\newcommand{\beml}{\begin{subequations}}
\newcommand{\eml}{\end{subequations}}
\newcommand{\beq}{\begin{eqnarray}}
\newcommand{\eq}{\end{eqnarray}}
\newcommand{\ba}{\begin{array}}
\newcommand{\ea}{\end{array}}
\newcommand{\bpm}{\begin{pmatrix}}
\newcommand{\epm}{\end{pmatrix}}
\newcommand{\bc}{\begin{cases}}
\newcommand{\ec}{\end{cases}}

\begin{document}
\title{Quantum magnetization fluctuations via spin shot noise}

\author{Alireza Qaiumzadeh}
\affiliation{Center for Quantum Spintronics, Department of Physics, Norwegian University of Science and Technology, NO-7491 Trondheim, Norway}

\author{Arne Brataas}
\affiliation{Center for Quantum Spintronics, Department of Physics, Norwegian University of Science and Technology, NO-7491 Trondheim, Norway}

\begin{abstract}
Recent measurements in current-driven spin valves demonstrate magnetization fluctuations that deviate from semiclassical predictions. We posit that the origin of this deviation is spin shot noise. On this basis, our theory predicts that magnetization fluctuations asymmetrically increase in biased junctions irrespective of the current direction. At low temperatures, the fluctuations are proportional to the bias, but at different rates for opposite current directions. Quantum effects control fluctuations even at higher temperatures. Our results are in semiquantitative agreement with recent experiments and are in contradiction to semiclassical theories of spin-transfer torque.
\end{abstract}

\date{\today}
\maketitle
Spin-transfer torque (STT), the transfer of spin angular momentum from spin-polarized currents to localized magnetic moments, is a cornerstone in spintronics \cite{STT1, STT2, STT3, STT4, Arne}. The low power consumption and scalable architecture open a promising path for STT-based devices in future data storage and information-processing technologies. For example, STT facilitates state-of-the-art nonvolatile random-access memory \cite{Arne, STTRAM,Racetrack} and spin-transfer nano-oscillators \cite{STNO}.

STT affects the magnetization dynamics via (an) a (anti)dampinglike torque whose amplitude is proportional to the current density. In the simplest manifestation, the magnetization dissipation is enhanced (antidamping torque) or reduced (damping torque) and is captured by an effective Gilbert damping constant.  At finite temperatures, there are additional random torques that fluctuate. In a semiclassical picture, stochastic temperature-dependent random magnetic fields model the fluctuations. These fields obey the dissipation-fluctuation theorem; the two-point autocorrelation function is proportional to the effective Gilbert damping parameter \cite{STT5, STT6}.

Recently, Zholud \textit{et al.} measured an anomalous behavior in the magnetization fluctuations in a current-driven spin valve \cite{Urazhdin}. Measurements at low temperatures suppress thermal fluctuations, but there were observations of magnetization fluctuations irrespective of the current direction. These measurements cannot be explained by semiclassical STT models \cite{STT1,STT2}. Rather, Ref. \cite{Urazhdin} suggests that quantum effects are essential. Thus far, there have only been a few theoretical works describing STT in a quantum picture \cite{Urazhdin,Sham}.
\begin{figure}[h]
\includegraphics[width=\columnwidth]{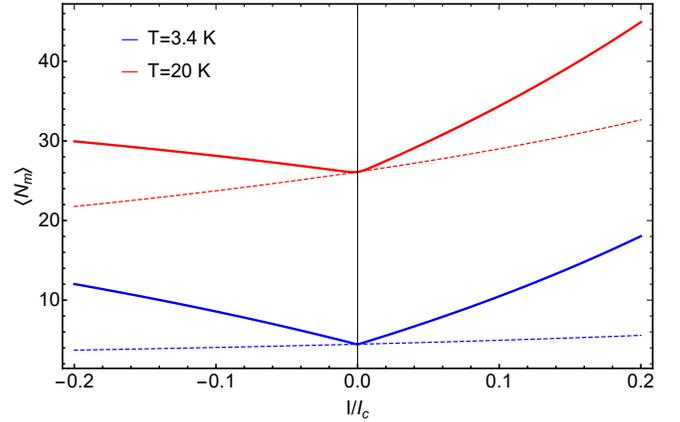}
\caption{(Color online) The spin fluctuations, which are proportional to angular-dependent magnetoresistance, as a function of dimensionless applied current at different temperatures. Solid lines and dashed lines show the results with and without the contribution of the quantum spin shot noise, respectively; see Eq. (\ref{magdensity}).}
\label{QSTT}
\end{figure}

In this Rapid Communication, charge \cite{Blanter,Nazarov}. Similarly, the discrete nature of itinerant electron spin in units of $\hbar/2$ causes spin shot noise \cite{Foros}.  We demonstrate that spin shot noise can explain the recent experimental observations. At low temperatures and when there is a current flowing through the spin valve, spin shot noise dominates and affects the distribution of spin fluctuations. These bias-driven quantum fluctuations exert additional stochastic torques on the magnetization dynamics that do not obey the fluctuation-dissipation theorem. As a result of the competition between the fluctuations and the dissipation (through the Gilbert damping), the magnons are driven out of equilibrium. Consequently, our theory predicts that the quantum magnetization fluctuations in spin valves depend on the bias voltage amplitude at low temperatures.

Our main finding is a microscopic expression for the total spin fluctuations in spin valves,
\begin{align}\label{main-result}
\langle N_m \rangle \simeq \frac{1}{1-\frac{I}{I_c}}\left[ f_\mathrm{BE}(\Omega, T) + \frac{1}{2}+\Xi^{(\mathrm{sh})}(U, T)\right],
\end{align}
where $I<I_c$ is the applied current, $I_c$ is the threshold switching spin-polarized current in the ferromagnetic layer \cite{STT4}, $f_\mathrm{BE}$ is the Bose-Einstein distribution function at temperature $T$ and magnon frequency $\Omega$, and $U$ is the applied bias voltage. In Eq.\ (\ref{main-result}), the prefactor is due to the semiclassical STT. The first term within the bracket is the contribution of thermal magnons. The second term arises from the vacuum fluctuations \cite{Landau}. Our main contribution is the third term that originates from quantum spin shot noise.  We will demonstrate that $\Xi^{(\mathrm{sh})}(U, T)$ is an even function of the bias voltage $U$.

As an illustration, in  Fig.\ \ref{QSTT}, we plot the spin fluctuations as a function of bias voltage for different temperatures. The curve is in semi-quantitative agreement with recent experimental results \cite{Urazhdin}. Moreover, we can quantify the shot noise contribution in terms of one parameter that we relate to the microscopic details of the system. We give a good estimate for this quantity. Beyond the scope of our work, we encourage detailed \textit{ab initio} evaluations of the shot noise parameter to explore further the quantitative consistency between our approach and measurements.

\begin{figure}[t]
\includegraphics[width=\columnwidth]{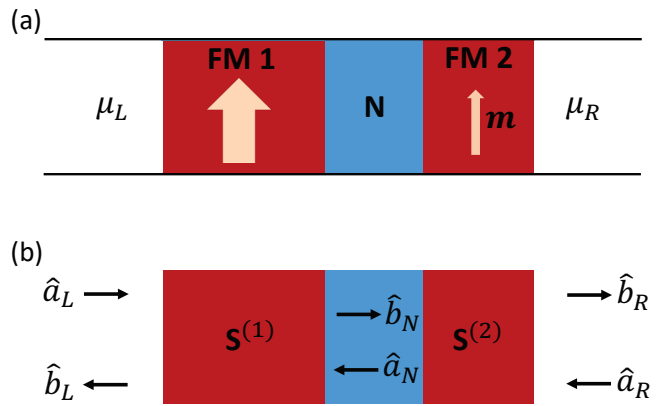}
\caption{(a) A spin-valve structure attached to left (L) and right (R) reservoirs with chemical potentials $\mu_L$ and $\mu_R$, respectively. FM1 is a hard ferromagnet, FM2 is a soft ferromagnet, and N is a normal metal. (b) A two-terminal scattering representation of the spin-valve structure in (a) with two spin-dependent scattering matrices $S^{(1)}$ and $S^{(2)}$. $\hat{a}_{A}$ and $\hat{b}_{A}$ are fermion annihilation operators representing incoming and outgoing electrons, respectively, in the left lead ($A=L$), middle of the normal metal ($A=N$), and the right lead ($A=R$).}
\label{SpinValve}
\end{figure}

To explain our approach, we first review how Zholud \textit{et al.} \cite{Urazhdin} measured the magnetic fluctuations of Eq.\ (\ref{main-result}). The experimental observation is the angular dependency of magnetoresistance in a spin-valve nanopillar, shown in Fig. \ref{SpinValve}, as a function of current. In spin valves, the anisotropic magnetoresistance depends on the relative angle $\theta$ between the magnetization directions in the free layer, FM2, and the polarizer, FM1 \cite{Gerrit}. The resistance varies as $R(\theta)=R_0+\Delta R (1-\cos\theta)/2$, where $\Delta R$ is the magnetoresistance amplitude. Meanwhile, the number of spin fluctuations is $\langle N_m\rangle=(M_\mathrm{s} \mathcal{V}/2\mu_B) (1-\cos\theta)$, where $M_\mathrm{s}$ is the saturation magnetization and $\mu_B$ is the Bohr magneton. Thus, there is a linear relation between  the spin fluctuations and the angular-dependent magnetoresistance: $\langle N_m\rangle=(M_\mathrm{s} \mathcal{V}/\mu_B)(R(\theta)- R(0))/\Delta R$. Therefore, measuring the magnetoresistance reflects the total magnetization fluctuations. Reference\ \cite{Urazhdin} finds an asymmetric enhancement of the magnetic fluctuations as a function of current, which is in contradiction to semiclassical STT theory.

To model the aforementioned experiment, we consider the spin-valve structure depicted in Fig. \ref{SpinValve}. This system consists of a hard ferromagnetic layer, FM1; a normal metal spacer, N; and a free ferromagnetic layer, FM2, attached to two left and right reservoirs, as shown in Fig. \ref{SpinValve}. We will derive Eq. (\ref{main-result}) and discuss its consequences. At finite temperature and when there are spin currents, the stochastic Landau-Lifshitz-Gilbert (LLG) equation describes the dynamics of the magnetization direction $\bm{m}$ in the free layer \cite{STT1, STT2, Brown},
\be \label{LLG}
\dot{\bm{m}}=-\gamma \bm{m} \times [\mathcal{\bm{H}}_{\mathrm{eff}}+\bm{h}(t)]+\alpha \bm{m}\times \dot{\bm{m}}+\beta \bm{m}\times (\bm{m} \times \hat{\bm{p}}),
\e
where $\gamma$ is the effective gyromagnetic ratio; $\mathcal{\bm{H}}_{\mathrm{eff}}=-(M_\mathrm{s} \mathcal{V})^{-1}\delta \mathcal{F}/\delta \bm{m}$ is the effective magnetic field, which is a functional derivative of the thermodynamic free energy $\mathcal{F}$ with respect to the magnetization;  $\bm{h}(t)$ is the stochastic magnetic field arising from various sources of fluctuations; $\alpha$ is the effective Gilbert damping parameter, $\beta=\gamma \hbar p I/(M_\mathrm{s} \mathcal{V} e)$ is the STT parameter, with $\hbar$ is the reduced Planck constant, and $p$ and $\hat{\bm{p}}$ parametrize the spin-current polarization and its spin direction, respectively; and $e$ is the electron charge. In the LLG equation [Eq. (\ref{LLG})], the first term on the right-hand side describes the magnetization precession around the local effective magnetic field, while the second term introduces a dissipative mechanism, the so-called Gilbert damping torque, which slows down the precession and pushes the magnetization towards the effective magnetic field. The third term is the (anti)dampinglike STT or Slonczewski spin-torque. An additional fieldlike spin torque is also allowed in the LLG equation but this torque is typically negligible in metallic spin valves and we disregard it \cite{STT4, Arne}.

There are two contributions to the stochastic field $\bm{h}$. First, there are intrinsic contributions related to the intrinsic Gilbert damping. According to the fluctuation-dissipation theorem, a stochastic and uncorrelated thermal field describes these effects. Our focus is on the second, extrinsic, contributions to the random field. These fields relate to the fluctuations of the spin transfer torque, the difference between the spin currents to the left and the right of the free magnetic layer. The spin-transfer torque fluctuations consist of equilibrium thermal fluctuations arising from the spin-pumping-induced enhancement of the Gilbert damping that just renormalized the first intrinsic contribution, and bias-driven quantum spin shot noise fluctuations.

We will now compute the stochastic field due to the spin-transfer torque fluctuations. On both sides of the soft ferromagnet (FM2), we define currents with respect to the flow towards the ferromagnet. The rate of change of the magnetization direction, the STT, is then  $-\gamma \hat{\bm{I}}_{\bm{\sigma}, \mathrm{abs}}(t)/(M_\mathrm{s} \mathcal{V})$, where the absorbed spin current is, $\hat{\bm{I}}_{\bm{\sigma}, \mathrm{abs}}(t)=\sum_{A\in\{N,R\}}\hat{\bm{I}}_{\bm{\sigma}, A}(t)$. Since the spin-transfer torque is transverse to the magnetization,  the associated stochastic magnetic field appearing in the LLG equation is given by $\bm{h}(t)=-(M_\mathrm{s} \mathcal{V})^{-1} \bm{m}\times \delta \hat{\bm{I}}_{\bm{\sigma}, \mathrm{abs}}(t)$, where $\delta \hat{\bm{I}}_{\bm{\sigma}, \mathrm{abs}}(t)$ is the deviation of the absorbed current from its average value.
The extrinsic fluctuating fields vanish on average $\langle h_i(t) \rangle=0$, and the transverse components are correlated as \cite{Foros}
\begin{align}
\langle h_i(t) h_j (t')\rangle&=-\frac{1}{M^2_\mathrm{s} \mathcal{V}^2}\sum_{A,B\in\{N,R\}} \mathcal{C}_{ji,AB}(t-t'),\nonumber\\
\langle h_i(t) h_i (t')\rangle&=\frac{1}{M^2_\mathrm{s} \mathcal{V}^2}\sum_{A,B\in\{N,R\}} \mathcal{C}_{jj,AB}(t-t'),\nonumber\\
\label{Spincurrent-corr1}
\end{align}
where $i\neq j=x,y$, $\mathcal{C}_{ij,AB}(t-t')=\langle \delta I_{\sigma_i,A}(t) \delta I_{\sigma_j, B}(t')\rangle$ is the spin-current correlation function and $\delta I_{\sigma_i}^{A}$ is the deviation of the vector component $i\in\{x, y, z\}$ of the spin current in the lead $A$, $\delta I_{\sigma_i,A}(t)=I_{\sigma_i,A}(t)-\langle I_{\sigma_i,A}(t)\rangle$.

The spin current is  $\mathbf{\hat{I}}_{\bm{\sigma},A}=-(\hbar/2e)\sum_{\alpha,\beta}\bm{\sigma}^{\alpha\beta}{\hat{I}}^{\beta\alpha}_{A}$, where $\bm\sigma$ is the vector of Pauli matrices. In a scattering formalism, the components of the spin current tensor are
\begin{align}
\hat{I}^{\alpha\beta}_{A}(t)&=\frac{e}{\hbar}\left(\hat{b}^\dag_{A,\alpha}(t)\hat{b}_{A,\beta}(t)-\hat{a}^\dag_{A,\alpha}(t)\hat{a}_{A,\beta}(t)\right),
\end{align}
where $\hat{a}_{A\alpha}$ and $\hat{b}_{A\alpha}$ are vector operators containing all transverse transport channels, which annihilate electrons with spin $\alpha$ in lead $A$ that move toward and away from the ferromagnets, respectively. For simplicity, in expressing the formulas, we drop all channel indices, but contributions from all channels are taken into account in all of our calculations and results.

Since our system consists of two ferromagnets, we need to explicitly compute the currents to the left and right of the free layer, FM2. To this end, we must relate the scattering properties in the subsystems. The outgoing and incoming modes are, in the absence of spin-flip processes \cite{spin-flip}, related via
\begin{align}\label{scattering-matrix1}
\left(
  \begin{array}{c}
    \hat{b}_{L\alpha} \\
    \hat{b}_{N\alpha} \\
  \end{array}
\right)
&=
\left(
  \begin{array}{cc}
    s^{(1)}_{LL\alpha} & s^{(1)}_{LN\alpha} \\
    s^{(1)}_{NL\alpha} & s^{(1)}_{NN\alpha} \\
  \end{array}
\right)
\left(
  \begin{array}{c}
    \hat{a}_{L\alpha} \\
    \hat{a}_{N\alpha} \\
  \end{array}
\right),\\ \label{scattering-matrix2}
\left(
  \begin{array}{c}
    \hat{a}_{N\alpha} \\
    \hat{b}_{R\alpha} \\
  \end{array}
\right)
&=
\left(
  \begin{array}{cc}
    s^{(2)}_{NN\alpha} & s^{(2)}_{NR\alpha} \\
    s^{(2)}_{RN\alpha} & s^{(2)}_{RR\alpha} \\
  \end{array}
\right)
\left(
  \begin{array}{c}
    \hat{b}_{N\alpha} \\
    \hat{a}_{R\alpha} \\
  \end{array}
\right).
\end{align}
The diagonal and off-diagonal elements of the scattering matrices represent the reflection and the transmission coefficients, respectively.  A scattering matrix that relates the annihilation operators associated with incoming and outgoing waves from the left and right leads is
\begin{align}
\label{scattering-matrix3}
\left(
  \begin{array}{c}
    \hat{b}_{L\alpha} \\
    \hat{b}_{R\alpha} \\
  \end{array}
\right)
&=
\left(
  \begin{array}{cc}
    s_{LL\alpha} & s_{LR\alpha} \\
    s_{RL\alpha} & s_{RR\alpha} \\
  \end{array}
\right)
\left(
  \begin{array}{c}
    \hat{a}_{L\alpha} \\
    \hat{a}_{R\alpha} \\
  \end{array}
\right).
\end{align}
The scattering matrix of the total system $S$ is related to the scattering matrices of FM1, $S^{(1)}$, and FM2, $S^{(2)}$, via \cite{Datta},
\begin{subequations}\label{scattering-amps}
\begin{align}
s_{RL\alpha} &= s^{(1)}_{NL\alpha}  s^{(2)}_{RN\alpha} \mathcal{D}_{\alpha},\\
s_{LR\alpha} &= s^{(2)}_{NR\alpha}  s^{(1)}_{LN\alpha} \mathcal{D}_{\alpha},\\
s_{LL\alpha} &= s^{(1)}_{LL\alpha}+s^{(2)}_{NR\alpha}  s^{(1)}_{LN\alpha} s^{(1)}_{NL\alpha} \mathcal{D}_{\alpha},\\
s_{RR\alpha} &= s^{(2)}_{RR\alpha}+s^{(1)}_{NN\alpha}  s^{(2)}_{RN\alpha} s^{(2)}_{NR\alpha} \mathcal{D}_{\alpha},
\end{align}
\end{subequations}
where $\mathcal{D}_{\alpha}=[1-s^{(1)}_{NN\alpha} s^{(2)}_{NN\alpha}]^{-1}$.

In order to compute the spin currents, we must express the currents in Eq. (\ref{Spincurrent-corr1}) as functions of the properties of the incoming modes in the left and right lead only, $\hat{a}_L$ and $\hat{a}_R$. Using Eqs. (\ref{scattering-matrix1})-(\ref{scattering-amps}), we can rewrite the operators in the normal spacer as a linear combination of the left and right leads as
\begin{subequations}
\begin{align}
\hat{a}_{N\alpha}=K^a_{L\alpha} \hat{a}_{L\alpha}+K^a_{R\alpha} \hat{a}_{R\alpha},\\
\hat{b}_{N\alpha}=K^b_{L\alpha} \hat{a}_{L\alpha}+K^b_{R\alpha} \hat{a}_{R\alpha},
\end{align}
\end{subequations}
where
\begin{subequations}
\begin{align}
K^a_{L\alpha}&=s^{(2)}_{NN\alpha} s^{(1)}_{NL\alpha}\mathcal{D}_{\alpha},\\
K^a_{R\alpha}&=s^{(2)}_{NR\alpha}\mathcal{D}_{\alpha},\\
K^b_{L\alpha}&=s^{(1)}_{NL\alpha}\mathcal{D}_{\alpha},\\
K^b_{R\alpha}&=s^{(1)}_{NN\alpha} s^{(2)}_{NR\alpha}\mathcal{D}_{\alpha}.
\end{align}
\end{subequations}
Using fermionic statistics and performing straightforward calculations, we can find the total correlator of the transverse components of the stochastic magnetic field as a function of the scattering matrix elements.
In the simplest limit, considering that the frequency of the spin-current noise is the lowest energy scale in the system, we can approximate $\mathcal{C}_{ij,AB}(t-t')\simeq \mathcal{C}_{ij,AB}(\omega=0)\delta(t-t')$.
The correlation function at zero frequency is given by,
\begin{align}
&\mathcal{C}_{ij,AB}(\omega=0)=\frac{\hbar}{8\pi}\sum_{\alpha\beta\in\{\uparrow,\downarrow\}}\sigma_i^{\alpha\beta}\sigma_j^{\beta\alpha} \delta_{AB} \nonumber\\ &\times\int d\varepsilon \left[\big(\mathcal{G}_1(\varepsilon)+\mathcal{G}_2(\varepsilon) \big)\delta_{AR}+\mathcal{G}_3(\varepsilon) \delta_{AN} \right], \label{Spincurrent-corr2}
\end{align}
where
\begin{subequations}\label{Spincurrent-corr22}
\begin{align}
\mathcal{G}_1=&\mathrm{Tr}\left[2\delta_{AB}-s^\dag_{AB\beta}(\varepsilon) s_{AB\alpha}(\varepsilon)-s^\dag_{BA\alpha}(\varepsilon) s_{BA\beta}(\varepsilon)\right] \nonumber\\&\times f_A(\varepsilon)(1-f_A(\varepsilon)),\\
\mathcal{G}_2=& \sum_{C,D\in\{L,R\}} \mathrm{Tr}[s^\dag_{AC\alpha} s_{AD\beta} s^\dag_{BD\beta} s_{BC\alpha}]\nonumber\\&\times f_C(\varepsilon)(1-f_D(\varepsilon)),\\
\mathcal{G}_3=&\sum_{C, D\in\{L,R\}}\sum_{l,m\in\{a,b\}}\mathrm{Tr}\big[(K^{l}_{C\alpha})^\dag K^{l}_{D\beta} (K^{m}_{D\beta})^\dag K^{m}_{C\alpha} \big] \nonumber\\&\times f_C(\varepsilon)(1-f_D(\varepsilon)) \, .
\end{align}
\end{subequations}
In Eqs.\ ({\ref{Spincurrent-corr22}}), the trace is a sum over the transverse channels and $f_{A}(\varepsilon)$ is the Fermi-Dirac distribution at energy $\varepsilon$ in lead $A$ with chemical potential $\mu_{A}$.

The correlator of Eq.\ (\ref{Spincurrent-corr2}) can be further simplified. Typically, the thermal energy $k_\mathrm{B} T$, and the applied bias potential $e U\equiv \mu_L-\mu_R$, are much smaller than the Fermi energy of reservoirs $\varepsilon_F$. Thus the scattering matrix elements can then be evaluated at the Fermi level. In this limit, we find that the total correlator of spin currents, Eq.\ (\ref{Spincurrent-corr2}) can be decomposed into two parts. The first part is the thermal contribution that vanishes at zero temperature, and the second part is the shot noise contribution, which is bias dependent and finite even at zero temperature. Finally, by using Eqs. (\ref{Spincurrent-corr1}), (\ref{Spincurrent-corr2}) and (\ref{Spincurrent-corr22}), the extrinsic stochastic thermal spin-current noise and spin shot noise correlators become
\begin{align}
&\langle h^{(\mathrm{th})}_i(t) h^{(\mathrm{th})}_j (t')\rangle=\xi^{(\mathrm{th})}\delta_{ij}\delta(t-t'),\label{Correlators1}\\
&\langle h^{(\mathrm{sh})}_i(t) h^{(\mathrm{sh})}_j (t')\rangle=\xi^{(\mathrm{sh})}\delta_{ij}\delta(t-t'),\label{Correlators11}
\end{align}
with the following correlator amplitudes,
\begin{align}
\xi^{(\mathrm{th})}(\omega=0)&=\frac{4\pi \alpha_{\mathrm{sp}}}{\gamma M_\mathrm{s} \mathcal{V}}k_\mathrm{B} T,\label{correlators2}\\
\xi^{(\mathrm{sh})}(\omega=0)&=\frac{\hbar  W_{RL}}{4\pi M^2_\mathrm{s} \mathcal{V}^2} \left(\frac{e U}{\tanh(\frac{e U}{2 k_\mathrm{B} T})}-2 k_\mathrm{B} T\right). \label{correlators3}
\end{align}

To derive the above results, we have used the following integrals: $\int d \varepsilon (f_L-f_R)^2=e U \coth(e U/2 k_\mathrm{B} T)-2k_\mathrm{B} T$ and $\int d \varepsilon f_{L(R)}(1-f_{L(R)})=k_\mathrm{B} T$. $\alpha_{\mathrm{sp}}$ is a Gilbert-type damping parameter arising from the spin pumping that depends on the spin-mixing conductance \cite{sp}. $W_{RL}$ is a function of the spin-dependent scattering matrices evaluated at the Fermi level,
\begin{align}
W_{RL}&=\mathrm{Tr}[s_{RL\uparrow}s^\dag_{RL\uparrow} s_{RR\downarrow} s^\dag_{RR\downarrow}]_{\varepsilon=\varepsilon_\mathrm{F}}\nonumber\\&+\sum_{l,m\in\{a,b\}}\mathrm{Tr}\big[K^{m}_{L\uparrow} (K^{l}_{L\uparrow})^\dag K^{l}_{R\downarrow} (K^{m}_{R\downarrow})^\dag \big]_{\varepsilon=\varepsilon_\mathrm{F}}, \label{W_LR}
\end{align}
The expression for $W_{LR}$ in Eq. (\ref{W_LR}) relates the strength of the shot noise contribution of the magnetization fluctuations to the microscopic features of the system. Hence, it is possible to compute $W_{LR}$ by first-principles calculations. Such extensive computations are beyond the scope of this work. Nevertheless, we can give good estimates for $W_{LR}$ in a similar manner that good estimates for the mixing conductance can be given without carrying out detailed \textit{ab initio} calculations. In the  Stoner model, we find that $W_{\mathrm{RL}}\approx \mathcal{N}\delta\varepsilon/\varepsilon_\mathrm{F}$, where $\mathcal{N}$ is the number of transverse modes and $\delta\varepsilon$ is the exchange splitting in the free ferromagnetic layer \cite{Foros}.

The extrinsic thermal spin-current noise of Eq. (\ref{correlators2}) is proportional to the spin-pumping-induced enhancement of the damping parameter $\alpha_{\mathrm{sp}}$  \cite{sp} and obeys the fluctuation-dissipation theorem. This term is analogous to the intrinsic contribution of the thermal noise arising from the intrinsic Gilbert damping $\alpha_0$. Thus, we can take the latter contribution into account by replacing $\alpha_{\mathrm{sp}}\rightarrow \alpha=\alpha_{\mathrm{sp}}+\alpha_0$, in Eq. (\ref{correlators2}). At finite frequencies, we could rewrite the total thermal noise correlator in the frequency domain as
\begin{subequations}\label{Correlators-th1}
\begin{align}
&\langle h^{(\mathrm{th})}_i(\omega) h^{(\mathrm{th})}_j (\omega')\rangle=\xi^{(\mathrm{th})}(\omega)\delta_{ij}\delta(\omega-\omega'),\\
&\xi^{(\mathrm{th})}(\omega)=\frac{2\pi \alpha \hbar\omega}{\gamma M_\mathrm{s} \mathcal{V} \tanh (\frac{\hbar\omega}{2 k_\mathrm{B} T})}.
\end{align}
\end{subequations}

Now, we discuss the shot noise correlator of Eq.\ (\ref{correlators3}). When the applied bias potential is larger than the thermal energy, $k_\mathrm{B} T\ll |e U| \ll \varepsilon_F$, the shot noise amplitude is $\xi^{(\mathrm{sh})} \propto |e U|$, whereas in the opposite limit, $|e U|\ll k_{\mathrm{B}}T\ll \varepsilon_F$, we obtain $\xi^{(\mathrm{sh})} \propto e^2 U^2/(6 k_{\mathrm{B}}T)$.

Finally, we calculate the total magnetization fluctuations in the presence of the spin shot noise as well as thermal stochastic noise [Eqs. (\ref{Correlators11}), (\ref{correlators3}) and (\ref{Correlators-th1})] for a uniaxial and collinear ferromagnetic layer. The total free energy of the free ferromagnetic layer consists of the anisotropy energy and the Zeeman energy, $\mathcal{F}=\mathcal{V}^{-1}\int d{\bm{r}} \left(-K (\bm{m}\cdot \hat{z})^2/2-\mu_B\bm{m}\cdot \bm{B}\right)$, where $K>0$ is the uniaxial anisotropy energy and $\bm{B}=B \hat{z}$ is the external magnetic field along the $z$-direction.  We expand the unit vector along the magnetization in terms of the transverse excitations $ \delta \bm{m}$, as
$\bm{m}=\sqrt{1-\delta \bm{m}^2}\hat{z}+\delta \bm{m}$, with $\hat{z}\cdot\delta \bm{m}=0$. The number of spin fluctuations is proportional to the small deviation of the magnetization along the equilibrium $z$-direction,  $\langle N_m\rangle=(M_\mathrm{s} \mathcal{V}/4\mu_B)\langle \delta \bm{m}^2\rangle$. Linearizing the LLG equation [Eq. (\ref{LLG})] in the presence of spin currents with a polarization in the $z$-direction results in an effective equation of motion for spin fluctuations
\begin{align}\label{magEq1}
i(1+i\alpha)\dot{\psi}(t)-\left(\Omega+i\beta\right)\psi(t)=\gamma\tilde{h}(t),
\end{align}
where $\psi(t)=\delta m_x(t)-i \delta m_y(t)$, $\tilde{h}(t)=h_x(t)-ih_y(t)$, and $\Omega=\gamma(K+\mu_B B)/(M_\mathrm{s} \mathcal{V})$ is the ferromagnetic resonance frequency. Through a Fourier transformation, the solution of Eq. (\ref{magEq1}) becomes
\begin{align}\label{magEq2}
\psi(\omega)=\frac{\gamma\tilde{h}(\omega)}{\omega-\Omega+i(\alpha \omega-\beta)}.
\end{align}
To obtain Eq.\ (\ref{main-result}), we compute the spin fluctuations in the limit of small damping and bias voltage as
\begin{align}\label{magdensity}
\langle N_m \rangle&=\frac{M_\mathrm{s} \mathcal{V}}{4\mu_B}\int\frac{d\omega}{2\pi} \frac{d\omega'}{2\pi} \langle \psi(\omega)\psi^{*}(\omega')\rangle\nonumber\\
&\simeq\frac{1}{1-\frac{I}{I_c}}\left[f_\mathrm{BE}(\Omega,T)+\frac{1}{2}+\frac{\gamma M_\mathrm{s}\mathcal{V}}{4\pi \hbar \alpha \Omega}\xi^{(\mathrm{sh})}\right],
\end{align}
where the threshold switching current is given by $I_c=\alpha\Omega M_\mathrm{s} \mathcal{V} e/(\gamma \hbar p)$. The total spin fluctuation has three contributions: The first term in Eq. (\ref{magdensity}) is the contribution of thermal magnons that obey the Bose-Einstein statistics; the second term is quantum zero-point fluctuations arising from the uncertainly in the ground state of spin components \cite{Landau}; and the third term is the contribution of the spin shot noise, which has a purely quantum mechanical nature and is finite even at zero temperature.

Figure \ref{QSTT}, shows the total spin fluctuation number of Eq. (\ref{magdensity}) as a function of the charge current for different temperatures. We consider a ferromagnetic thin-film layer of permalloy in the presence of a magnetic field of $1.5$T with a ferromagnetic resonance of $\Omega=100$ GHz and an effective Gilbert damping of $\alpha=0.01$. There is a zero-bias singularity in the magnetization fluctuations at zero temperature [see Eq. (\ref{magdensity})] that is rapidly broadened by increasing the temperature, see Fig. \ref{QSTT}. This broadening is not due to the contribution of thermal magnons but is rather the contribution of spin shot noise at finite temperature. The piecewise and asymmetric dependence of the magnetization fluctuations to the bias current survives even at higher temperatures. In Fig. \ref{QSTT}, we also compare the magnon fluctuations with and without the contribution from the quantum shot noise. In the absence of quantum shot noise (dotted lines), depending on the direction of the applied bias voltage, the magnetization fluctuations increase due to the antidampinglike STT or decrease because of the dampinglike STT. The quantum spin shot noise, on the other hand, leads to an increase in the magnetization fluctuations irrespective of the current direction.

Zholud \textit{et al.} \cite{Urazhdin} introduce a phenomenological model to describe the effects of localized spin fluctuations on spin transfer. At zero temperatures, they find $\langle N(I) \rangle\approx[(|I|+pI)/(2pI_c)]/(1-I/I_c)$. In contrast, we microscopically compute that quantum spin shot noise carried by the itinerant electrons significantly contributes to the spin fluctuations. At zero temperatures, Eq. (\ref{magdensity}) becomes $\langle N_m \rangle\approx (G_0 R W_{LR}/8\pi)(|I|/2pI_c)/(1-I/I_c)$, where $G_0=2e^2/h$ is the conductance quantum and $R=V/I$ is the spin-valve resistance. We suggest carrying out experiments with different spin polarizations of the injected current into the free layer to distinguish between the two contributions.

In summary, in addition to the semiclassical picture of STT \cite{STT1, STT2}, there is an important and so far overlooked quantum effect arising from the spin shot noise contribution. This effect originates from the discrete nature of itinerant electron spins. At low temperatures, the resulting quantum fluctuations strongly affect the total magnetization fluctuations in spin valves. The result is in good agreement with the recent observation of a piecewise-linear dependence of the quantum magnetization fluctuation on the applied current measured by Zhould \textit{et al}.

\section*{Acknowledgments}
The research leading to these results was supported by the European Research Council via Advanced Grant No. 669442, ``Insulatronics,'' and by the Research Council of Norway through its Centres of Excellence funding scheme, Project No. 262633, ``QuSpin.''

\textit{Note added---} Recently, we became aware of another paper \cite{Scott} that attributes the quantum STT \cite{Urazhdin} to the spin fluctuations of magnetic junctions.


\begin{thebibliography}{9}

\bibitem {STT1}
J. C. Slonczewski, J. Magn. Magn. Mater. \textbf{159}, L1 (1996).

\bibitem {STT2}
L. Berger, Phys. Rev. B \textbf{54}, 9353 (1996).

\bibitem {STT3}
J. A. Katine, F. J. Albert, R. A. Buhrman, E. B. Myers, and D. C. Ralph, Phys. Rev. Lett. \textbf{84}, 3149 (2000).

\bibitem {STT4}
D. C. Ralph and M. D. Stiles,  J. Magn. Magn. Mater. \textbf{320}, 1190 (2008).

\bibitem {Arne}
A. Brataas, A. D. Kent, and H. Ohno, Nat. Mater. \textbf{11}, 372 (2012).


\bibitem {STTRAM}
H. Zhao,  A. Lyle,  Y. Zhang,  P. K. Amiri,  G. Rowlands,  Z. Zeng,  J. Katine,  H. Jiang,  K. Galatsi,  K. L. Wang,  I. N. Krivorotov, and  J.-P. Wang, J. Appl. Phys. \textbf{109}, 07C720 (2008).

\bibitem {Racetrack}
S. S. P. Parkin and M. Hayashi, L. Thomas, Science \textbf{320}, 190 (2008); S. Parkin, S. H. Yang, Nat. Nanotechnol. \textbf{10}, 195 (2015).

\bibitem {STNO}
J.A. Katine, E. E. Fullerton, J. Magn. Magn. Mater. \textbf{320}, 1217 (2008).


\bibitem {STT5}
Z. Li and S. Zhang,  Phys. Rev. B \textbf{69}, 134416 (2004).

\bibitem {STT6}
R. H. Koch, J. A. Katine, and J. Z. Sun,  Phys. Rev. Lett. \textbf{92}, 088302 (2004).

\bibitem {Urazhdin}
A. Zholud, R. Freeman, R. Cao, A. Srivastava, and S. Urazhdin, Phys. Rev. Lett. \textbf{119}, 257201 (2017).

\bibitem {Sham}
Y. Wang and L. J. Sham, Phys. Rev. B \textbf{85}, 092403 (2012); \textbf{87}, 174433 (2013); Y. Wang, W.-q. Chen, and F.-C. Zhang, New. J. Phys. \textbf{17}, 053012 (2015).

\bibitem {Blanter}
Y. M. Blanter and M. B\"{u}ttiker, Phys. Rep. \textbf{336}, 1 (2000).

\bibitem {Nazarov}
\textit{Quantum Noise in Mesoscopic Physics}, edited by Y. V. Nazarov (Kluwer, Dordrecht, 2003).


\bibitem {Foros}
J. Foros, A. Brataas, Y. Tserkovnyak, and G. E. W. Bauer, Phys. Rev. Lett. \textbf{95}, 016601 (2005); J. Foros, A. Brataas, G. E. W. Bauer, and Y. Tserkovnyak, Phys. Rev. B \textbf{79}, 214407 (2009).

\bibitem {Landau}
L. D. Landau and E. M. Lifshitz, \textit{Statistical Physics} (Pergamon, New York, 1980), Part I.

\bibitem {Gerrit}
G. E. W. Bauer, Y. Tserkovnyak, D. Huertas-Hernando, and A. Brataas, Phys. Rev. B \textbf{67}, 094421 (2003); A. Shpiro, P. M. Levy, and S. Zhang, Phys. Rev. B \textbf{67}, 104430 (2003).

\bibitem {Brown}
W. F. Brown, Jr., Phys. Rev. \textbf{130}, 1677 (1963).

\bibitem{spin-flip}
F. J. Jedema, M. S. Nijboer, A. T. Filip,, and B. J. van Wees, Phys. Rev. B \textbf{67}, 085319 (2003).

\bibitem {Datta}
S. Datta, \textit{Electronic transport in mesoscopic systems} (Cambridge University Press, Cambridge, U.K., 1999).

\bibitem{sp}
Y. Tserkovnyak, A. Brataas, and G. E. W. Bauer, Phys. Rev. Lett. \textbf{88}, 117601 (2002).


\bibitem {Scott}
S. A. Bender, R. A. Duine, and Y. Tserkovnyak, arXiv: 1808.00777.


\end{thebibliography}
\end{document}